\documentclass[twocolumn]{aastex62}

\received{}
\revised{}
\accepted{}
\submitjournal{ApJ}

\shorttitle{}
\shortauthors{Annibali et al.}

\begin{document}

\title{HST resolves stars in a tiny body falling on the dwarf galaxy DDO~68}

\correspondingauthor{Francesca Annibali}
\email{francesca.annibali@inaf.it}

\author{F. Annibali}
\affil{INAF- Astrophysics and Space Science Observatory \\
 Via Piero Gobetti, 93/3, 40129 - Bologna, Italy}

\author{M. Bellazzini}
\affiliation{INAF- Astrophysics and Space Science Observatory \\
 Via Piero Gobetti, 93/3, 40129 - Bologna, Italy}

\author{M. Correnti}
\affiliation{Space Telescope Science Institute \\
3700 San Martin Drive, Baltimore, MD 21218, USA}

\author{E. Sacchi}
\affiliation{Space Telescope Science Institute \\
3700 San Martin Drive, Baltimore, MD 21218, USA}

\author{M. Tosi}
\affiliation{INAF- Astrophysics and Space Science Observatory \\
 Via Piero Gobetti, 93/3, 40129 - Bologna, Italy}

\author{M. Cignoni}
\affiliation{Dipartimento di Fisica, Universit\`a di Pisa,\\
Largo Bruno Pontecorvo, 3, I-56127 Pisa, Italy}
\affiliation{INFN, Sezione di Pisa,\\
Largo Bruno Pontecorvo, 3, I-56127 Pisa, Italy}

\author{A. Aloisi}
\affiliation{Space Telescope Science Institute \\
3700 San Martin Drive, Baltimore, MD 21218, USA}

\author{D. Calzetti}
\affiliation{Department of Astronomy, University of Massachusetts-Amherst \\
710 N Pleasant Street, Amherst, MA 01003-9305, USA}

\author{L. Ciotti}
\affiliation{Dipartimento di Fisica e Astronomia, Universit\`a di Bologna,\\
Via Gobetti 93/2, I-40129 Bologna, Italy}

\author{F. Cusano}
\affiliation{INAF- Astrophysics and Space Science Observatory \\
 Via Piero Gobetti, 93/3, 40129 - Bologna, Italy}

\author{J. Lee}
\affiliation{Caltech-IPAC \\
1200 E California Blvd, Pasadena, CA 91125}

\author{C. Nipoti}
\affiliation{Dipartimento di Fisica e Astronomia, Universit\`a di Bologna,\\
Via Gobetti 93/2, I-40129 Bologna, Italy}



\begin{abstract}

We present new Hubble Space Telescope (HST) imaging of a stream-like system associated with the dwarf galaxy DDO~68, located in the Lynx-Cancer Void at a  distance of D$\sim$12.65 Mpc from us. 
The stream, previously identified in deep Large Binocular Telescope images as a diffuse low surface brightness structure, is resolved into 
individual stars in the F606W (broad V) and F814W ($\sim$I) images acquired with the Wide Field Camera 3. The resulting V, I color-magnitude diagram (CMD) of the resolved stars is dominated by old (age$\gtrsim$1-2 Gyr) red giant 
branch (RGB) stars. From the observed RGB tip, we conclude that the stream is at the same distance as DDO~68, confirming the physical association with it. 
A synthetic CMD analysis indicates that the large majority of the star formation activity in the stream occurred at epochs earlier than $\sim$1 Gyr ago, and that the star formation at epochs more recent than $\sim$500 Myr ago is compatible with zero. The total stellar mass of the stream is $\sim10^{6} M_{\odot}$, about 1/100 of that of DDO~68. This is a striking example of hierarchical merging in action at the dwarf galaxy scales.

\end{abstract}

\keywords{galaxies: dwarf galaxies --- 
galaxies: irregular galaxies ---
galaxies: low surface brightness galaxies ---
galaxies: interacting galaxies ---
stellar astronomy: stellar populations}


\section{Introduction} \label{sec:intro}

The dwarf galaxy DDO~68, with a stellar mass of $\sim10^8 \ M_{\odot}$  \citep{Sacchi16} and a dynamical mass of 
$M_{dyn} \gtrsim 5.2 \times 10^9 M_{\odot}$ \citep{Cannon14}, is a rare observed example of merger occurring at the scale of dwarf galaxies. Located at a distance of $\sim$12.65 Mpc from us \citep{Sacchi16}, it resides in a very empty region of space,  in the Lynx-Cancer Void \citep{Pustilnik11}; nevertheless, it has been suggested to be in the process of accreting (or interacting with) one or more smaller companions. 
\cite{Ekta08} hypothesised a late-stage merger between two gas-rich progenitors to explain the observed distortions in the HI. 
Using Hubble Space Telescope (HST) data, \cite{Tikhonov14} proposed that DDO~68 is in fact composed of two distinct systems, a main component, 
DDO~68~A, and a disrupted satellite (dubbed DDO~68~B) being currently accreted by DDO~68~A and producing the ``cometary tail''. \cite{Cannon14} detected a more remote, low-mass ($M_{HI}\sim3\times 10^7 M_{\odot}$) companion, located at a projected distance of $\sim$40 kpc from DDO~68 and connected to the galaxy through a bridge of low surface brightness HI gas ($N_{HI}\sim 2 \times 10^{18} cm^{-2}$). Indeed, the occurrence of mergers and interaction events could help explain some of the peculiarities observed in DDO~68, such as 
 its very distorted morphology, characterized by the presence of a ``cometary tail'', the unusual distribution of its H~II regions, mostly located in the head and tail of the ``comet''  and almost absent in the innermost galaxy regions \citep{Moiseev14,Pustilnik17}, and its very low metallicity \citep[$\sim$2\% solar, e.g.][which is one of the lowest metallicities ever measured in a star-forming galaxy]{Pustilnik05,Izotov09,Annibali19}, much too low for its mass. For instance, \cite{lee04} showed that the inclusion of interacting galaxies can increase the scatter in the luminosity-metallicity relation and may force the observed correlation toward lower metallicities and/or larger luminosities/masses.

\begin{figure}
\plotone{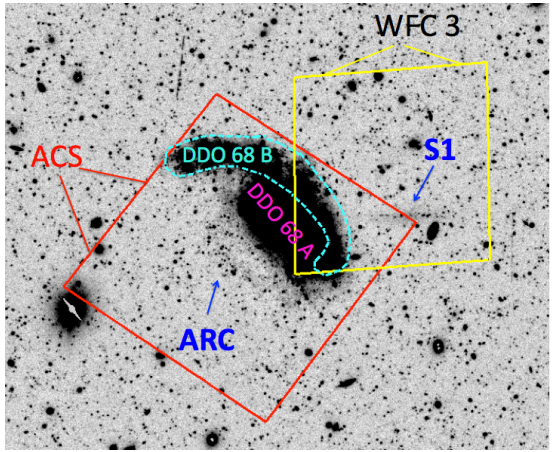}
\caption{500$\times$400 arcsec$^2$ portion of our LBT LBC image of DDO~68 showing the arc and the stream S1 identified by \cite{Annibali16}. The dashed cyan contour indicates the approximate division of DDO~68 into the A and B sub-systems, as proposed by \cite{Tikhonov14}. Overplotted on the LBT image are the footprints of our previously acquired ACS data \citep{Sacchi16} and of the new WFC3 data.   \label {fig:image_all}}
\end{figure}

\begin{figure*}
\plotone{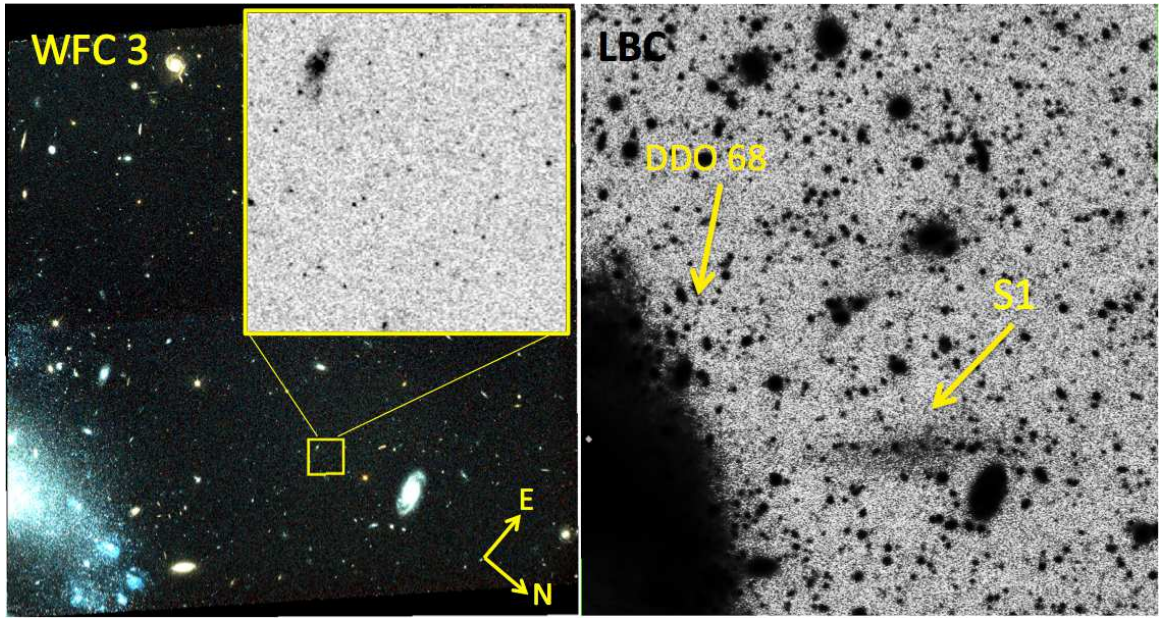}
\caption{Left: HST WFC3 color-combined image (F606W$=$blue, F814W$=$red) centered on the faint stream accreting onto DDO~68. 
The stream S1 is not identifiable in the HST images due to its low surface brightness, but its stars are very well resolved.
The insertion provides a  zoom-in view of a 10$\times$10 arcsec$^2$ ($\sim$600$\times$600 pc$^2$) field in F606W showing individual stars resolved in S1. 
Right: the LBT LBC  image in g where we identified for the first time the faint stream \citep{Annibali16}.  \label {fig:image}}
\end{figure*}

In \cite{Sacchi16}, we derived the  star formation history (SFH) of DDO~68 using HST Advanced Camera for Surveys  (ACS) data in F606W ($\sim$ broad V) and F814W ($\sim$I). We found that DDO~68, despite its extremely low metallicity, hosts a conspicuous population of red giant branch (RGB) stars, implying that the galaxy has been forming stars since at least 1$-$2 Gyr ago, and possibly since epochs as early as a Hubble time. 
The galaxy star formation activity  has been fairly continuous over all the look-back time, with an enhancement by factor of $\gtrsim$2  between 10 and 100 Myr ago
(at a rate of $\sim 4 \times 10^{-2} \ M_{\odot} \ yr^{-1} \  kpc^{-2}$), and a lower activity at current epochs. 
Interestingly, when studying the spatial distribution of stars of different ages, we recognized  a peculiar concentration of RGB stars toward an edge of the ACS chip, as also identified by \cite{Tikhonov14}. We noticed that the magnitudes of these stars were compatible with DDO~68' s distance but,  due to the poor statistics, we could not draw firm conclusions.

With the aim of getting insights into the origin of these RGB stars, and into DDO~68' s merging history more in general,
we exploited the powerful combination of large field of view  ($\sim$23$\times$23 arcmin$^2$) and high photometric
depth provided by the Large Binocular Cameras (LBC) on the Large Binocular Telescope (LBT),  and acquired new deep $g$- and $r$- imaging of DDO~68. The LBT data showed that the peculiar concentration of RGB stars detected in the ACS images are in fact part of a more extended low surface brightness  (${\rm \mu_r\sim 28.7 \ mag \ arcsec^{-2}}$) stream-like system connected to DDO~68, which we dubbed S1 \citep{Annibali16}. 
Fig.~\ref{fig:image_all} displays a portion of our LBT image with indicated the stream S1 and an identified low surface brightness arc.
S1 has a projected size of $\sim$20$\times$80 arcsec$^2$ in the LBT images, a total integrated magnitude of $m_r = 21.6 \pm 0.4$ mag, and an average color of $g-r = 0.56 \pm 0.11$, which translate, under the assumption of the same DDO~68' s distance of $\sim$12.65 Mpc, into a physical size of $\sim$1.2 $\times$ 5 kpc$^{2}$ and a stellar mass in the range $(1.5 - 6)\times 10^5 M_{\odot}$. In \cite{Annibali16} we showed through N-body simulations that while DDO~68's ``cometary tail'' can be 
reproduced by the accretion of a ten times less massive body, S1 requires the accretion of a third body with mass of about one hundredth that of DDO~68.

In this paper we present new HST WFC3/UVIS data acquired during Cycle 24 with the main 
purposes of resolving the entire  S1 system into individual stars, improving the statistics with respect to our previous ACS data, and 
deriving a conclusive distance for S1 via the RGB tip method. Probing the physical association of S1 with DDO~68 through its distance is a fundamental step to confirm a rare observed case of multiple accretions onto an isolated  dwarf galaxy as small as DDO~68. 

Indeed, N-body simulations in a  $\Lambda$ Cold Dark Matter ($\Lambda$CDM) 
cosmology predict that structures are assembled through a continuous hierarchical 
merging process, and a consequence is that present-day galaxies, both giants and dwarfs, are expected to be surrounded and orbited 
by a large number of smaller systems or satellites  \citep[e.g.][]{Diemand08}. However, while there is ample observational evidence for this to occur  
around giant galaxies  (e.g., in the Milky Way \citep{Belokurov06}, in Andromeda \citep{Ibata01,McConnachie09}, in several Local Volume spirals \citep{Martinez10} and also giant ellipticals such as Centaurus A \citep{CenA}), little observational evidence of accretion events into LMC-size galaxies has been reported in the literature so far: among these rare examples are the LMC itself, with several convincing satellites \citep{Koposov18}, and the magellanic irregular galaxy NGC~4449, with evidence for a stellar stream \citep{Martinez12,Rich12} and for another disrupted dwarf companion \citep{Annibali12}. However, accretion phenomena onto galaxies less massive than $\sim 10^9 M_{\odot}$ in stars have remained elusive so far \citep{Amorisco14,Kacharov17,Cicu18}, although panoramic imaging surveys have revealed possible candidate satellites of dwarf galaxies (the Antlia~B dwarf,  a likely satellite of the dwarf galaxy NGC~3109 \citep{sand15}, and Dw2, likely associated to the ultra-diffuse dwarf galaxy Dw1 in the CenA group \citep{crn19}) and very compact groups composed of only dwarfs 
 \citep{stierwalt17}. Therefore the  striking case of DDO~68 is a rare observed example of hierarchical merging in action at the dwarf galaxy scales 
predicted by the $\Lambda$ Cold Dark Matter scenario.

The paper is structured as follows. Observations and data reduction are described in Section~\ref{sec:obs}. 
The stream's color-magnitude diagram (CMD) is presented in Section~\ref{sec:cmd}, while we investigate the spatial distribution 
of the different stellar populations in Section~\ref{sec:star_map}. In Section~\ref{sec:distance} we provide evidence that S1 is located at the same 
distance of DDO~68 from us. In Section~\ref{sec:sfh} we derive S1's star formation history through the method of the synthetic color-magnitude diagrams. Finally, we present our discussion and conclusions  in Section~\ref{sec:discussion}.

\section{Observations and Data Reduction} \label{sec:obs}

The observations were performed in December 2017 with
the WFC3 UVIS channel in the F606W (broad V) and F814W (I) 
broad-band filters (GO program 14716, PI Annibali).
We preferred the WFC3 to the ACS because of the ageing of ACS and the more severe impact of 
charge transfer efficiency (CTE) effects compared to WFC3, and the F606W filter to the F555W filter because 
of the higher sensitivity of the former. 
This is the same filter setup adopted in our previous ACS program centered on DDO~68 (GO program 11578; PI Aloisi), 
allowing for a direct comparison between the photometry.  

 \begin{figure*}
\plotone{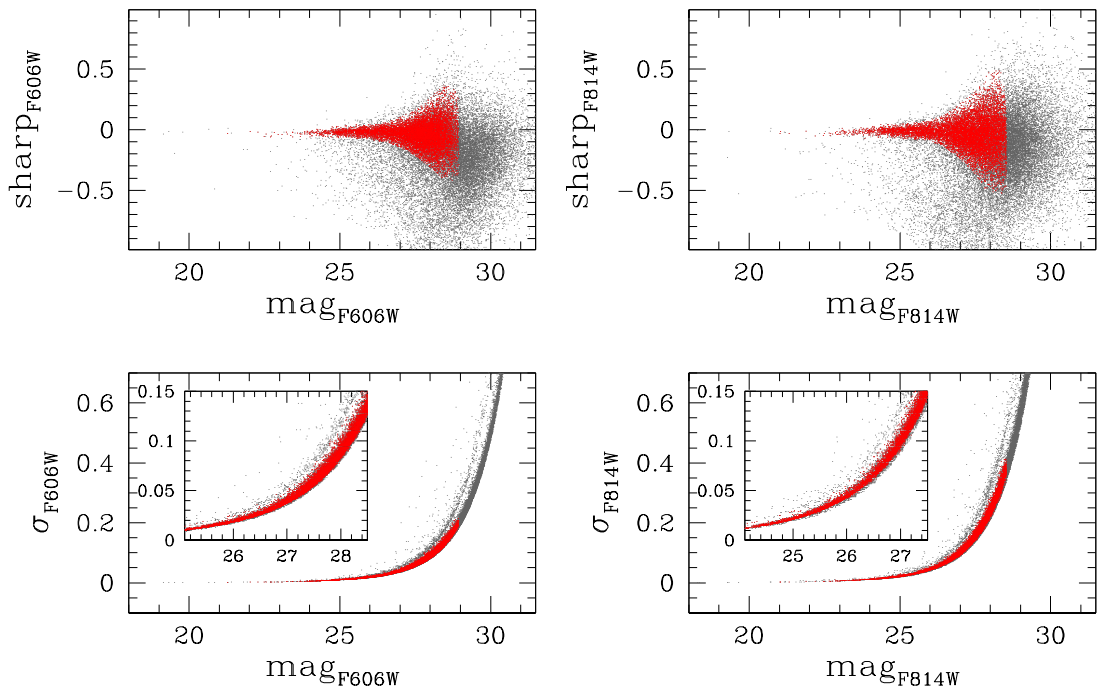}
\caption{Dolphot sharpness parameter versus magnitude  (top panels) and   magnitude error versus magnitude (bottom panels) 
in both F606W and F814W. Grey dots represent all sources selected adopting a Dolphot object-type flag of 1, as described in Section~\ref{sec:obs}, while red dots denote our adopted selections in the four planes. The two insertions in the bottom panels provide a zoom-in view of our error selection excluding sources in the ``upper'' sequences; these objects correspond to spurious detections on the WFC3 chip edges.  \label{fig:dolphot_par}}
\end{figure*}

The WFC3 UVIS camera was positioned with a specific orientation, with the gap between the two detectors rotated by $\sim$ 55 degrees NE, in order to place the whole S1 system within chip 2, thus avoiding the gap (see Fig.~\ref{fig:image}). 
A portion of the WFC3 field of view intentionally overlaps with our previous ACS observations of DDO~68. 
We observed S1 for total exposure times of $\sim$13,000 s in F606W and $\sim$13,000 s in F814W, implying a total of 10 HST orbits. The exposure times were determined assuming for S1 the same distance of DDO~68 \citep[$\sim$12.65 Mpc or $(m-M)_0\sim$30.51,][]{Sacchi16}, and 
requiring to reach down to 1 mag below the tip of the Red Giant Branch (i.e. down to I$\sim$27.4, V$\sim$28.7) with a signal-to-noise ratio of $\sim$5. 
For each filter, the five HST orbits were split into two separate visits lasting two and three orbits, respectively. Each visit was organized into eight sub-exposures, executed with a primary 2-point, $\sim$2.4 arcsec separation dither pattern to fill the gap between the WFC3 chips, combined with a secondary 4-point half $+$ integer pixel dither pattern to improve the PSF sampling, and get rid of cosmic rays and hot/bad pixels.   
Within the longer 3-orbit visit, we added two short $\sim$300 s exposures, executed with a 2-point spacing dither pattern ($\sim$4.8 arcsec separation), in order 
to get photometry of bright objects saturated in the long exposures.

We retrieved from the HST archive the {\texttt *.flc} science images, which are the  bias-corrected, dark-subtracted, flat-fielded, 
and CTE corrected data products from the CALWF3 pipeline \citep{wfc3_cal}. The individual {\texttt *.flc} images were then combined together into a single stacked, distortion-corrected reference image (i.e., the {\texttt *.drc} image) using the {\it Drizzlepac} software \citep{gonz+12}. 
More specifically, we first aligned all the images in the same filter using the software {\it TweakReg}, with the final transformations providing alignment of individual images to better than 0.02 pixels. Then, using the software {\it AstroDrizzle}, bad pixels and cosmic rays were flagged and rejected from the input images, and the input undistorted and aligned frames were combined together into a final stacked image. The final stacked images were generated at the native resolution of the WFC3/UVIS camera (i.e., 0.04 arcsec/pixel$^{-1}$).

To perform stellar photometry, we used the latest version of {\texttt Dolphot}\footnote{\url{http://americano.dolphinism.com}} \citep[][and numerous subsequent updates]{dolp00}. This software includes pre-processing and photometry modules customized  to each HST camera as well as precomputed point spred functions (PSFs) tailored
 to each filter. PSF fitting photometry was performed on the individual {\texttt *flc} images  using the combined drizzled {\texttt *drc} image as a reference frame. In order to achieve an optimal image alignment and source detection, the {\texttt Dolphot} parameters were set using a hybrid combination between the values recommended by the {\texttt Dolphot} manual and those adopted in \citet{will+14}.

The final output photometric catalog contains position and photometry of individual stars in F606W, F814W Vegamag, as well as several diagnostic parameters 
that can be used to exclude remaining artifacts and spurious detections. 
From the total catalog, we selected objects with a Dolphot ``object type'' flag of 1, corresponding to ``good stars'' that are well fitted by the PSF. 
However, background contaminants and spurious detections are likely to be still present in the catalog, and therefore we applied further cuts in the sharpness versus magnitude and error versus magnitude planes, as shown in Fig.~\ref{fig:dolphot_par}. Similar selections have proven to be very effective in removing a major fraction of contaminants from the photometric catalogs in our previous studies of dwarf galaxies \citep[e.g.,][]{Sacchi16}. We notice in particular that the error versus magnitude distributions in both F606W and F814W consist of two separate ``sequences'', with the ``upper'' one being only populated by spurious detection at the edges of the WFC3 chips.  

Confusion with background galaxies is a potential issue given the relatively low number of stars that we expect to resolve in S1; 
for this reason,  we eventually performed a visual inspection of the individual sources on the WFC3 color-combined V, I image to remove obvious background galaxies that had survived to our selection cuts. We caution however that very compact background galaxies, indistinguishable from individual stars, will 
still be present in the final catalog. Our final cleaned photometric catalog consists of $\sim$10,800 objects.

A portion of the WFC3 field of view overlaps part of the ACS images presented by \cite{Sacchi16}. For a consistency check between the ACS and WFC3 photometry, we cross-correlated our catalog with that of \cite{Sacchi16} and directly compared the F606W and F814W magnitudes for the stars in common between the two catalogs. Although the ACS and WFC3 photometries agree quite well within the errors, we notice the presence of a small systematic $\sim$0.03-0.04 offset in both filters between the two datasets, with the Sacchi et al. magnitudes being fainter than ours. 
We notice that this discrepancy can not be ascribed to photometric calibration differences between ACS and WFC3 \citep{acs_wfc3_cal}, and it is likely due to uncertainties in the applied aperture corrections.

\subsection{Artificial star tests} \label{sec:arti}

Artificial star tests were performed on the WFC3 images  in order to evaluate the incompleteness and the photometric errors as a function of magnitude. 
Indeed, the errors produced by photometric packages are typically underestimated \citep[e.g.][]{will+14} and do not provide a realistic evaluation of the photometric uncertainties. 
The adopted method, described in detail in previous papers \citep[e.g.,][]{Sacchi16,Cignoni16}, consists in adding artificial stars to the images and then in 
reducing the frames following the same procedure as for the real data.  Artificial stars were injected into the images following a uniform distribution; furthermore, they were added one at a time to not artificially alter the crowding on the images.  Color and magnitude intervals of the artificial stars were chosen to cover the observed CMD range and reaching down to $\sim$3 mag fainter than the observed magnitude limits. Each artificial star of given magnitude and color was then injected simultaneously into the F606W and F814W frames at one position in order to reproduce a realistic situation and to account for the correlation between the completeness in the two bands.  From the artificial star output catalog, a star was considered ``recovered'' if it had a measured magnitude within 0.75 mag from its input value in both filters, and if it satisfied all the selection cuts that were applied to the real photometry (see Section~\ref{sec:obs},  
Fig.~\ref{fig:dolphot_par}). 
Our tests provide a 50\% completeness at F606W$\sim$28 and F814W$\sim$27.5 in the crowded main body of DDO~68, and at F606W$\sim$28.5 and F814W$\sim$28 in the less crowded outer field. From the output minus input magnitude distributions, we estimate 1-$\sigma$ 
photometric errors of $\sim$0.06 and $\sim$0.13 at F606W$=$27 and 28, and of $\sim$0.16 and $\sim$0.32 at F814W$=$27 and 28.

\section{Color-Magnitude Diagram} \label{sec:cmd}

\begin{figure}
\plotone{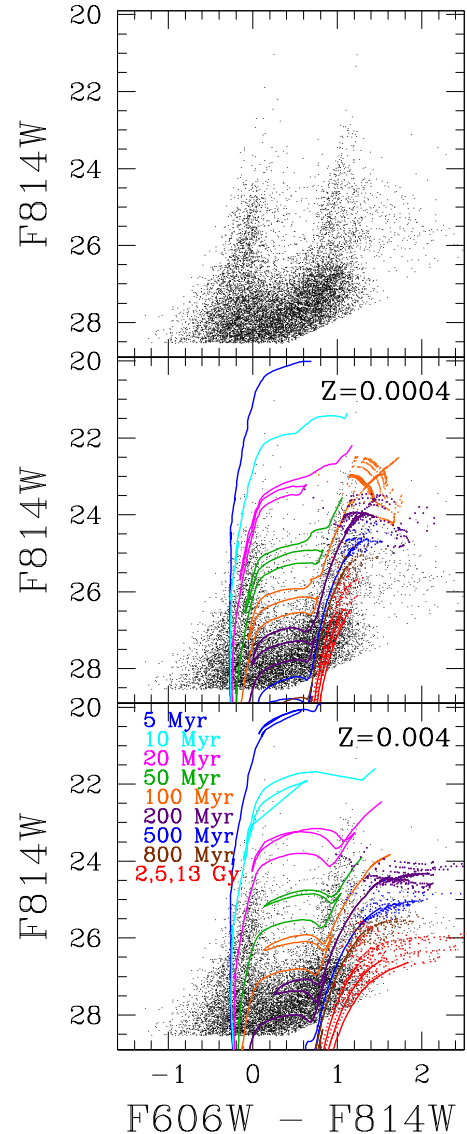}
\caption{I, V-I color-magnitude diagram for sources in our final cleaned photometric catalog (see Section~\ref{sec:obs} for details). Superimposed are the PARSEC stellar isochrones \citep{Bressan12,Marigo17} for  metallicities Z$=$0.0004 (middle panel) and Z$=$0.004 (bottom panel), and for different ages as indicated in the legend. The isochrones have been displayed assuming a distance modulus of $(m-M)_0=30.54$ (see Section~\ref{sec:distance}) and a foreground reddening of E(B$-$V)$=$0.016.\label{fig:cmd}}
\end{figure}

The total cleaned F814W vs. F606W$-$F814W CMD for the sources measured in the whole WFC3 field is shown in the top panel of Figure~\ref{fig:cmd},
while in the middle and bottom panels we compare the observed CMD with the PARSEC$+$COLIBRI stellar isochrones \citep{Bressan12,Marigo17} for two different metallicities: one of Z$=$0.0004 ($\sim$1/40 solar), consistent with the measured DDO~68' s H~II region oxygen abundance, and the other of Z$=$0.004, which is  more compatible with the metallicity expected from the galaxy  total luminosity. The displayed CMD was not corrected for reddening, while the isochrones were shifted assuming a distance modulus of $(m-M)_0=30.54$ (see Section~\ref{sec:distance}) and a foreground reddening of E(B$-$V)$=$0.016 \citep{Schlafly11};  
internal dust extinction within S1 is expected to be negligible  given the modest HI column densities at the position of S1 \citep{Cannon14} and 
the extremely low metallicity of DDO~68.

The CMD exhibits all the  evolutionary features typically observed in late-type dwarf galaxies: a well defined blue plume at F606W-F814W$\sim$-0.05 and F814W$\lesssim$26.5, populated by massive, young (age$\lesssim$10 Myr) main sequence stars and hot post-main sequence 
stars (age$\lesssim$50 Myr); 
a red plume at F606W-F814W$\sim$1 and F814W$\lesssim$26.5, containing red supergiants and  asymptotic giant branch (AGB) stars   
with ages from $\sim$10 Myr up to a few Gyrs; the red horizontal feature at  F814W$\sim$25.5 and 1$\lesssim$F606W-F814W$\lesssim$2.5 due to Carbon AGB stars in the thermal pulses phase (hereafter, TP-AGB), with ages from a few hundred Myrs up to a couple Gyrs; intermediate colour objects at 0$\lesssim$F606W-F814W$\lesssim$0.8, F814W$>$26.5,  which are intermediate-mass, core helium-burning stars with ages between $\sim$100 and $\sim$300 Myr (older stars in this evolutionary phase fall below our detection limit); finally, a concentration of objects at F814W$>$26.5 and 0.6$\lesssim$F606W-F814W$\lesssim$1.2, which is due to red giant branch (RGB) stars, and implies ages older than 1-2 Gyr and potentially as old as the age of the Universe. The same evolutionary features were observed by \cite{Sacchi16} for their CMD of the entire DDO~68 galaxy obtained from ACS data. 

\begin{figure*}
\plotone{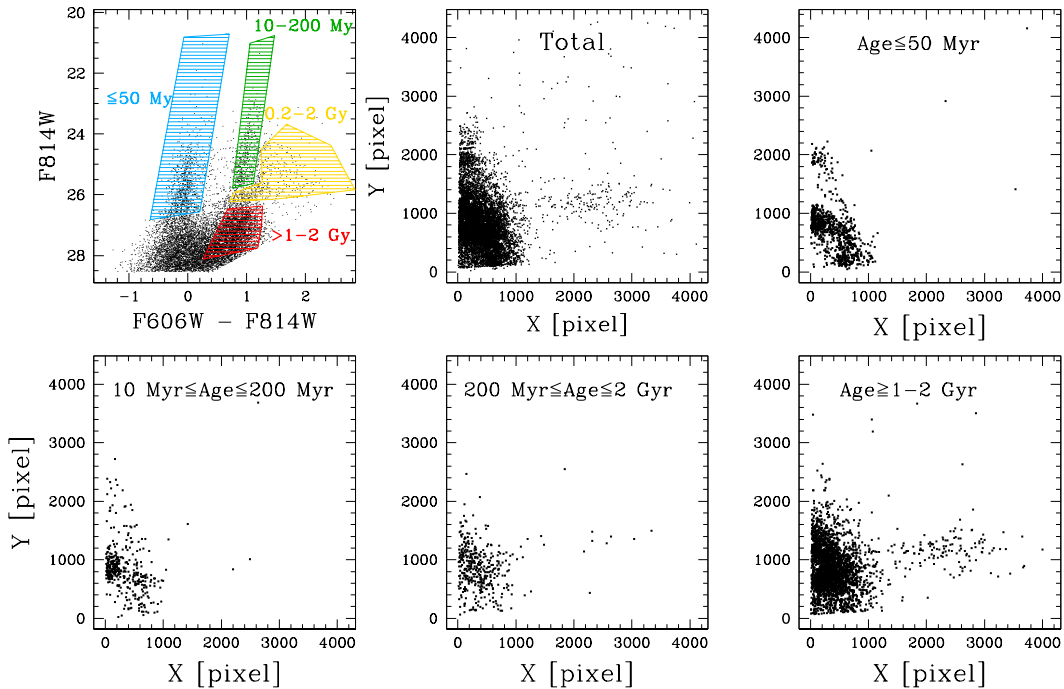}
\caption{Top left: color-magnitude diagram for sources in our final cleaned photometric catalog,  with overlaid selected regions corresponding 
to four age intervals, as indicated by the labels. Other panels: spatial distributions for the totality of sources in the catalog, and for stars with ages $\lesssim$50 Myr, in the ranges 10 Myr$-$200 Myr and 200 Myr$-$2 Gyr, and older than 1-2 Gyr. The concentration of objects at X$\lesssim$1000, Y$\lesssim$2000 is due to stars belonging to DDO~68, while the elongated feature at 1400$\lesssim$X$\lesssim$3400, 800$\lesssim$Y$\lesssim$1600 is the stream-like S1 system. \label{fig:spatial}}
\end{figure*}

Fig.~\ref{fig:cmd} shows that the overall appearance of the observed CMD is qualitatively well reproduced by the Z$=$0.0004 models, in agreement with the previous analysis of \cite{Sacchi16}.  In particular, the F606W$-$F814W color of the RGB is compatible with isochrones in the 2-13 Gyr age range, while models as metal-rich as Z$=$0.004 provide a poor match to this evolutionary phase unless ages of  $\lesssim$2 Gyr are assumed. On the other hand, the F606W$-$F814W color of younger stars 
(ages$\lesssim$100 Myr) in the blue and red plumes is less sensitive to the effect of metallicity, and strong claims concerning their metal content are not possible.  Finally, we notice that the Z$=$0.0004 models are not able to reproduce the observed red horizontal feature due to TP-AGB stars, which seems to better agree with models of higher metallicity. Indeed, DDO~68 is a strong outlier in the mass-metallicity relation, and its extremely low H~II region 
oxygen abundance has been suggested  to be due to the accretion of metal-poor gas; admittedly, we can not definitely exclude from the observed CMD the presence of a relatively young (age$\lesssim$1-2 Gyr) stellar population with metallicity several times higher than that observed in its H~II regions 
\citep[see also][]{Tikhonov14}.

\section{Resolved star maps}  \label{sec:star_map}

We investigated how stars in the total cleaned photometric catalog are distributed over the WFC3 field of view. Fig.~\ref{fig:spatial} shows spatial distributions for the totality of stars and for groups of stars belonging to different age intervals, as selected from the CMD. Guided by the stellar isochrones displayed in Fig.~\ref{fig:cmd}, we identified four regions in the CMD roughly corresponding to four different age intervals (see top left panel of Fig.~\ref{fig:spatial}): 
blue plume stars with ages $\lesssim$50 Myr; red plume stars, in the range $\sim$10 Myr$-$ $\sim$200 Myr; AGB stars with ages in the range 
$\sim$200 Myr$-$ $\sim$2 Gyr;  RGB stars, with ages older than 1$-$2 Gyr, and potentially as old as $\sim$13 Gyr.  
The spatial distribution of the totality of stars in the top central panel of  Fig.~\ref{fig:spatial} undoubtedly shows that the faint stream S1 identified in our previously acquired LBT images is resolved into individual stars in the WFC3 data. No obvious substructure other than S1 is identified in the resolved star maps. While the DDO~68 portion sampled by our WFC3 data contains stars of all ages, in agreement with the findings of \cite{Sacchi16},  S1 does not host young stars, and it is only populated by intermediate-age and old stars. 

 DDO~68 is also one of the 50 star-forming galaxies targeted by the 
LEGUS HST Treasury program \citep[ID 13364, PI Daniela Calzetti, see e.g.][]{Calzetti15}; hence, WFC3-UVIS photometry is available in the F275W, F336W and F438W bands in a field of view that partially covers also the stream. No bona-fide star can be identified in the stream in any of these short-wavelength images, in agreement with the lack of a young population in the F606W, F814W CMD. 
A more in-depth investigation of the S1's star formation history will be performed in Section~\ref{sec:sfh} through synthetic CMDs.

\subsection{Structural properties of S1}

\begin{figure}
\plotone{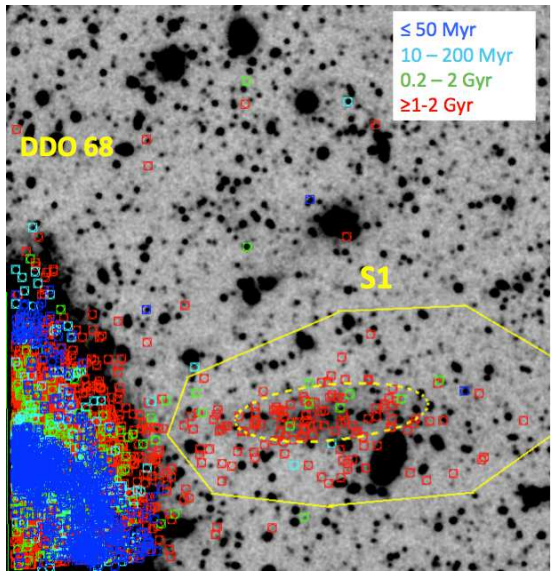}
\caption{Portion of the LBT/LBC image presented in \cite{Annibali16} with superimposed the stars resolved in the WFC3 data. 
The displayed image portion has the same field of view of our WFC3 images. 
Blue, cyan, green and red denote different age intervals, as selected from the CMD in Fig.~\ref{fig:spatial}: $\lesssim$50 Myr, 
in the ranges 10 Myr$-$200 Myr and 200 Myr$-$2 Gyr, and older than 1-2 Gyr. The yellow continuous polygin encloses  S1's extension as visually identified from the resolved star map, while the dashed contour is the fitted ellipse at major-axis half-light radius. \label{fig:lbt_res}}
\end{figure}

Stars belonging to different age intervals, as selected from the CMD in Fig.~\ref{fig:spatial}, are overplotted on the LBT LBC image in Fig.~\ref{fig:lbt_res} for a direct comparison. The resolved stars perfectly overlap with the unresolved low surface brightness feature identified in the LBT data. We can  
thus use the spatial distribution of AGB and RGB stars to infer some structural parameters of S1: the resolved star map is best fitted by ellipses with 
major axis oriented along a position angle of $PA=36^{+10}_{-18}$ deg (N to E), and minor over major axis ratio of b/a$\sim$0.29 (or ellipticity $\epsilon=1-b/a\simeq0.71$). Assuming star counts as a proxy for light, which is reasonable given the spatially uniform, mainly old population of S1, we estimate a major-axis half-light radius of $R_{e,maj}\simeq0.5'$, or $\simeq$1.9 kpc at S1' s distance (see Section~\ref{sec:distance}). The circularized effective radius is $R_{e,c} = R_{e,maj}\times \sqrt{b/a}\simeq0.27'$, or $\simeq$1 kpc. The fitted elliptical contour at major-axis half-light radius is shown in Fig.~\ref{fig:lbt_res}.

\section{Distance} \label{sec:distance}

\begin{figure*}
\plotone{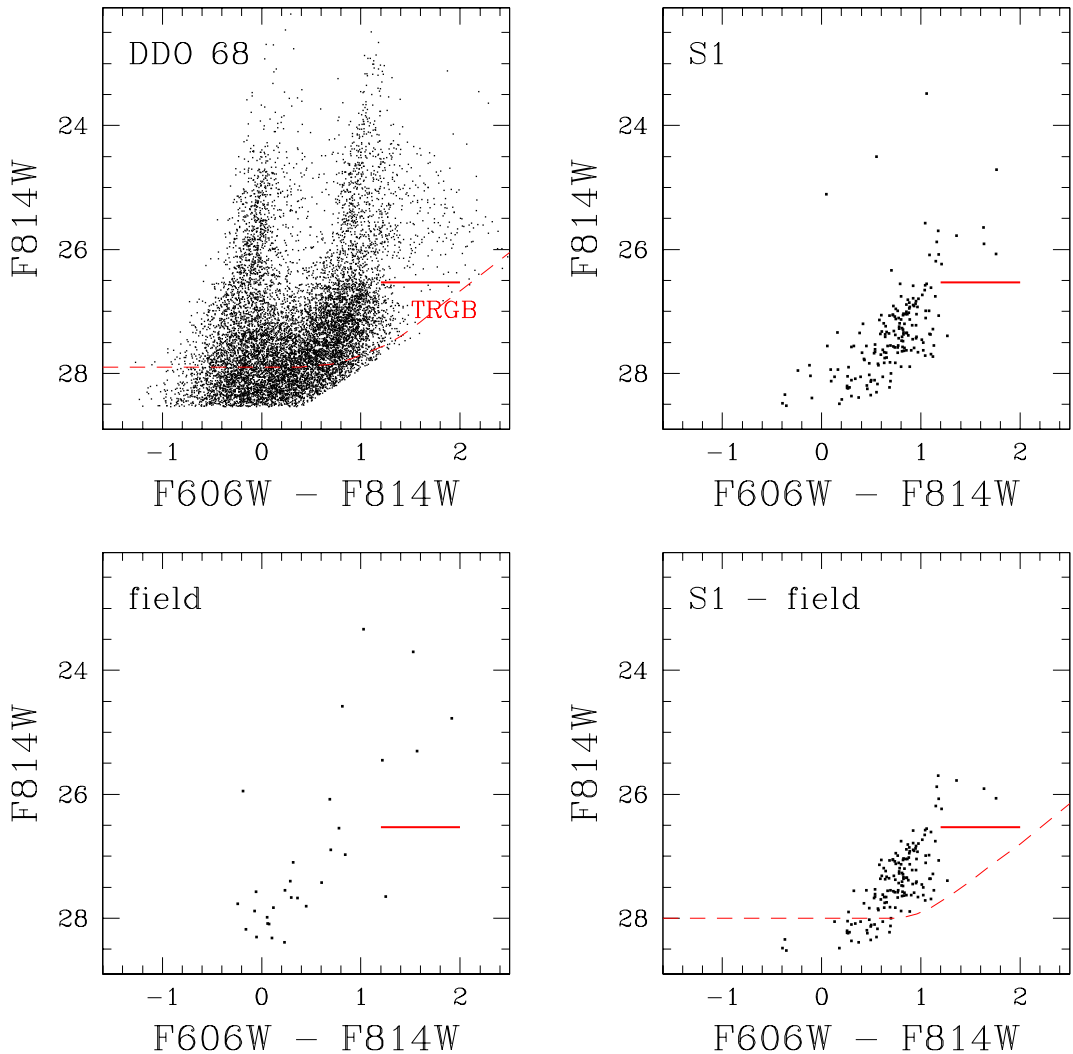}
\caption{CMDs for the portion of DDO~68 sampled by our WFC3 data (top left panel), for the S1 system, selected to be within the yellow polygon displayed in  Fig.~\ref{fig:lbt_res} (top right), for a control-field with the same S1's area (bottom left), and for S1 minus the contribution of the field (bottom right). 
The dashed curves indicate the 50\% completeness levels for DDO~68 and for S1.\label{fig:stream_cmd}}
\end{figure*}

In this section we use the tip of the RGB to test the hypothesis that S1's distance is compatible with that of DDO~68; 
this is a critical step to verify the scenario in which the two systems are physically associated and their observed 
connection is not a mere chance superposition. 
Fig.~\ref{fig:stream_cmd} shows the CMDs of the DDO~68 portion sampled by our WFC3 data, of S1, selected within the yellow polygon displayed in  Fig.~\ref{fig:lbt_res}, of a control-field with the same S1's area, and of S1 after subtraction of the field contribution; this was accomplished by ``merging'' the CMDs  of the field and of S1, and by removing from S1 the objects that were closer, in terms of color and magnitude, to the 
sources detected in the field. We notice that,  at magnitudes brighter than F814W$\lesssim$27.5  (i.e., $\sim$1 mag below the RGB tip), the field contamination to S1 can be considered negligible. 
There is a total of 154 stars in the S1 CMD after subtraction of the field contamination.

The red horizontal segment  shows the position of the tip of the RGB  as determined by applying a Sobel filter to the F814W luminosity function (LF) of stars with $0.6<F606W-F814W<1.2$ in DDO~68  \citep[following, e.g.,][]{Bellazzini11}.  
The RGB tip of DDO~68 is consistent with that visually identified in the S1's CMD, indicating that the two systems are qualitatively at the same distance. The low number of stars populating the RGB of S1 prevents a sensible quantitative determination of the tip position in this system \citep[see, e.g.,][for discussion]{Madore95,Bellazzini02}. Moreover, even in the case of well populated RGBs, the final uncertainty in the distance modulus is of order $\pm 0.1$~mag (see below), corresponding to $\simeq 0.5$~Mpc at the distance of DDO~68. This would not provide any additional information with respect to the direct comparison shown in Fig.~\ref{fig:stream_cmd}.
We will use synthetic CMDs in Section~\ref{sec:sfh} to provide further support to the agreement between DDO~68's and S1's distances. 

To derive DDO~68's distance, we convert the WFC3 F606W, F814W magnitudes into Johnson-Cousins V, I  applying the transformations of  \cite{Harris18}. 
We then apply a Sobel filter to the I LF and, after correcting for a foreground extinction of $A_I=0.028$ \citep{Schlafly11}, we obtain  $I_{TRGB,0}=26.49$.
If we input the RGB tip color,  $V-I_{TRGB,0}=1.04$, into Eq. 2 of \cite{Bellazzini08}, we obtain an expected absolute tip magnitude of $M_{I,TRGB}=-4.05$; this implies a distance modulus of $(m-M)_0=30.54 \pm 0.12$, corresponding to a distance of $D=12.8 \pm 0.7$ Mpc. This value is fully consistent 
within the errors with the distance of D$\sim$12.65 Mpc determined by \cite{Sacchi16}. At this distance, the projected distance between the center of DDO~68 and the center of the stream is $\sim$6 kpc.

\section{Star Formation History of S1} \label{sec:sfh}

The star formation history of S1 was derived through the synthetic CMD method 
\citep{Tosi91} using the Star Formation Evolution Recovery Algorithm 
(SFERA) \citep[see e.g.,][]{Cignoni15,Sacchi18}, which is a hybrid-genetic algorithm 
(genetic plus simulated annealing) based on a Poissonian approach.   
Synthetic CMDs were constructed starting from the PARSEC-COLIBRI 
stellar isochrones \citep{Bressan12,Marigo17}, assuming a Kroupa initial mass function 
\citep[IMF;][]{Kroupa01} from 0.1 to 300 M$_{\odot}$, a 30\% binary fraction\footnote{As shown by \cite{Cignoni16}, the assumed binary fraction does not have a significant impact on the SFH results.}, and 
metallicities in the range Z$=$0.0001$-$0.008 (i.e., from $\sim$1/150th solar to $\sim$half solar). 
In the color-magnitude plane, the isochrones were shifted assuming a distance modulus of $(m-M)_0=30.54$ 
(see Section~\ref{sec:distance}) and a foreground reddening of E(B$-$V)$=$0.016 \citep{Schlafly11}, while 
internal reddening within S1 is expected to be negligible given the lack of large amounts of gas at its position \citep{Cannon14} and the extremely 
low metallicity of DDO~68. 
The models were convolved with photometric errors and 
incompleteness as evaluated from artificial star tests (Section~\ref{sec:arti}).  
Synthetic CMDs were compared on a statistical way to the field-decontaminated CMD of S1 shown in the bottom right panel of Fig.~\ref{fig:stream_cmd}.   
The search for the solution was done considering only regions of the CMD above the 50\% completeness limit, translating into 
a magnitude limit of F814W$\sim$28. 

\begin{figure}
\plotone{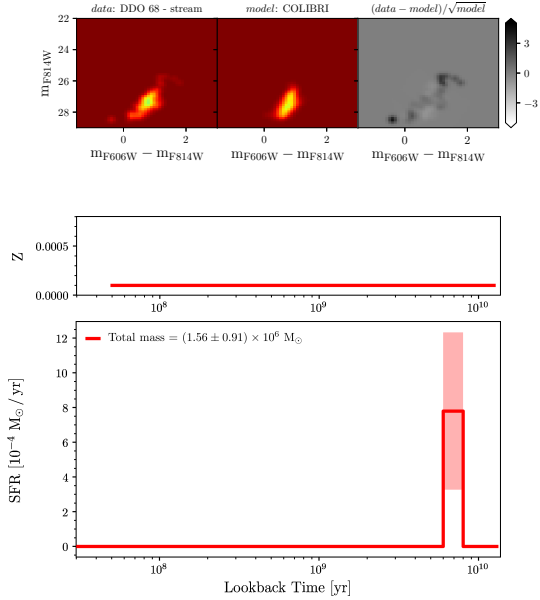}
\caption{Results from a simple stellar population (SSP) fit analysis applied to S1. The top panels display the  
Hess diagrams of the observed  CMD, of the best-fit SSP and of the residuals. The middle and bottom panels provide the metallicity  ($Z=10^{-4}$) and age 
(7$\pm$1 Gyr) of the best-fit SSP.  The total formed stellar mass is reported in the bottom panel.\label{fig:ssp}}
\end{figure}

Given the poor statistics  affecting the CMD of S1, we adopted in the first place the simplest possible approach and fitted the CMD with simple stellar populations (SSPs)  for a wide range of ages (from a few hundred Myr to 13 Gyr) and metallicities ($Z=10^{-4}-8\times10^{-3}$). From this analysis, the best agreement is provided by a 7$\pm$1 Gyr old SSP with metallicity of   $Z=10^{-4}$; the resulting stellar mass is 1.6$\pm$0.9$\times10^{6} M_{\odot}$. The comparison between the observed and the synthetic CMDs is shown in Fig.~\ref{fig:ssp}. We notice that, 
despite the  good match in the RGB  phase, the model fails in accounting for the observed  TP-AGB stars at magnitudes just brighter than the RGB tip.  
 Indeed, dwarf galaxies are known to host a quite complex and overly continuous SF, and a SSP fit may not provide an optimal description of the S1 evolution.  Therefore, we 
 performed a more sophisticated analysis based on the assumption a complex SFH, as described in the following.

To simulate a complex SFH, we adopted five age-bins  ($\leq$100 Myr, 100$-$500 Myr, 500 Myr$-$2 Gyr, 2$-$6 Gyr, and  6$-$13.7 Gyr), and let the metallicity freely vary in the range $Z=10^{-4}-8\times10^{-3}$, with the only constraint that it could not decrease by more than 25\% with respect to an older adjacent bin. The choice of just five coarse age bins is motivated by the low number of stars in the CMD, the large photometric errors, and the reached magnitude limit of just $\sim$1.5 mag below the RGB tip. Indeed, from the existence of an RGB we know that stars older than 1-2 Gyr are  present in S1, but we are not able to infer in great detail the SFH from the age of the Universe to $\sim$2 Gyr ago.

\begin{figure}
\plotone{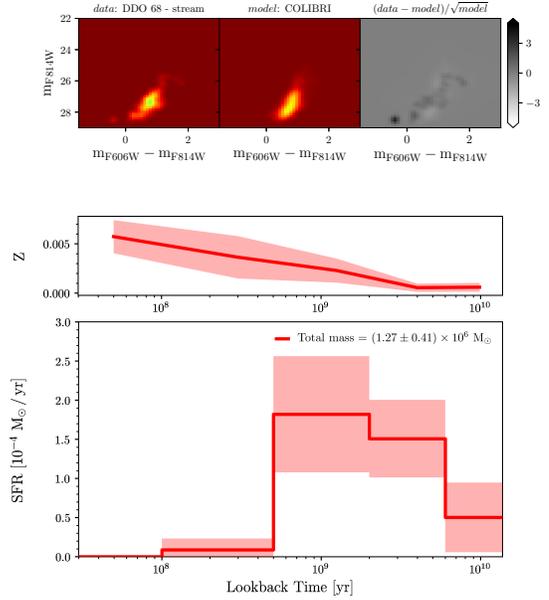}
\caption{Results of the SFERA procedure for the SFH recovery in S1. Top panels: 
Hess diagrams of the observed  CMD, of the best-fit model, and of the residuals. Middle panel: derived metallicity behaviour 
as a function of look-back time. Bottom panel: derived star formation rate as a function of look-back time. The total formed stellar mass is indicated.\label{fig:sfh1}}
\end{figure}

The best-fit solution from SFERA is presented in Fig.~\ref{fig:sfh1}; in the top panels, we show the empirical 
and the model CMDs in the form of Hess diagrams, i.e., density plots, and the residuals in units of Poisson uncertainties  ((data$-$model)/$\sqrt{model}$). As visible from the residuals, the model reproduces quite well the observed CMD above  F814W$\sim$28. The TP-AGB star region at F814W$\sim$26, F606W$-$F814W$\gtrsim$1.2 is better populated than in the previous SSP-fit case, although a shortage of counts is still present in the synthetic CMD. 
 However, we do not consider this partial discrepancy as a problem here, since it is known that this evolutionary phase is very difficult to model and still lacks proper calibration over the whole range of metallicity regimes.

The middle and bottom panels of Fig.~\ref{fig:sfh1} show the resulting metallicity and star formation behaviours as a function of look-back time. 
The solution recovered by SFERA implies that the vast majority of the SF activity occurred at epochs earlier than $\sim$500 Myr ago; the peak of the activity is predicted to have occurred in the 2$-$6 Gyr bin, while the adjacent  500 Myr$-$2 Gyr and 6$-$13.7 Gyr intervals were characterized by a lower SF.  As already expected from the CMD appearance in Fig.~\ref{fig:stream_cmd}, the SF rate in S1 was compatible with zero at epochs more recent than $\sim$0.5 Gyr ago.  
At earlier epochs we are not able to infer the details of the SFH with the data in our hands; however, the total formed stellar mass  of  1.27$\pm$0.41$\times10^{6} M_{\odot}$ can be considered a quite robust result and is in reasonable agreement with the estimates from both the SSP-fitting and the integrated light in \cite{Annibali16}.

Concerning metallicity, the best-fit solution implies a continuous increase in Z starting from $\sim$0.0006 at the oldest look-back times and reaching up to Z$\sim$0.006 ($\sim$40\% solar) at recent epochs. This value is significantly higher than the 
Z$\sim$0.0004 ($\sim$2\%) solar metallicity measured in DDO~68's H~II regions, and also much higher than expected for a $\sim10^{6} M_{\odot}$ stellar mass galaxy. Admittedly, given the strong age-metallicity degeneracy in the RGB phase combined with the relatively large photometric errors of our data, 
we can not claim a robust constraint on metallicity at epochs earlier than 1 Gyr ago. At more recent epochs, when the rate was compatible with zero, the metallicity 
is not well constrained either, but is very unlikely to have increased much. 
Notice that the reddening-metallicity degeneracy is not expected to play a significant role given the most likely low internal extinction within S1.

\begin{figure}
\plotone{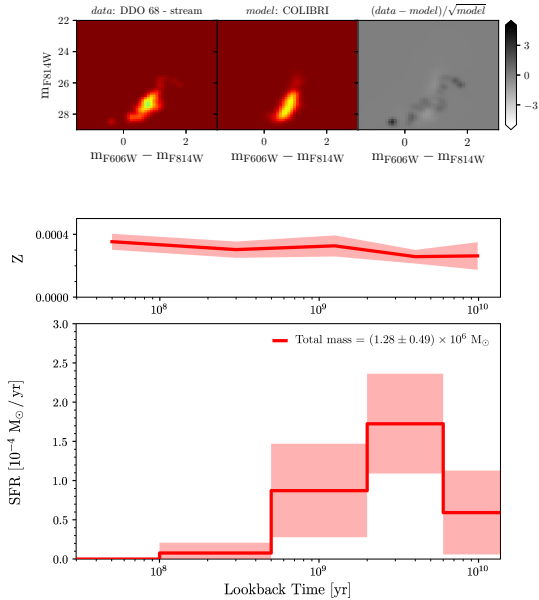}
\caption{Same as Fig.~\ref{fig:sfh1}, but with the assumption of a  Z$=$0.0004 metallicity upper limit. \label{fig:sfh2}}
\end{figure}

In order to test the effect of metallicity on our solution, we performed additional simulations imposing a Z$=$0.0004 metallicity upper limit. The results from this second run are shown in Fig.~\ref{fig:sfh2} following the same format of Fig.~\ref{fig:sfh1}. 
The total mass  of formed stars is 1.28$\pm$0.49$ \times10^{6} M_{\odot}$. 
The new solution still provides a satisfactory agreement between the observed and the synthetic CMDs, with the only difference of a worse match in the TP-AGB star region with respect to the previous run. Indeed, the presence of these stars may have driven SFERA toward the relatively metal-rich solution obtained in the first run. However, that the SFH and the total stellar mass obtained in the two different runs are in agreement with each other, within the uncertainties, indicates that these results are robust against metallicity uncertainties.

\section{Discussion and Conclusions} \label{sec:discussion}

We have presented new HST WFC3 imaging  in F606W (V) and F814W (I) of a low surface brightness stream-like system, dubbed S1, that is associated with the dwarf galaxy DDO~68 and that was identified for the first time in deep  LBT/LBC imaging. Previously acquired  HST ACS/WFC images centered on DDO~68 covered a small portion of S1, showing that it was resolved into individual stars, but lacking sufficient statistics to probe its distance and definitively  demonstrate its physical association with the dwarf galaxy. The new WFC3 data cover the entire S1 extension and provide a CMD sufficiently populated to assert that S1 is compatible with DDO~68's distance of 12.8$\pm$0.7 Mpc. 
This result rules out  the possibility of a chance superposition and demonstrates that S1 is indeed physically associated to DDO~68. 

While DDO~68 hosts stars of all ages, from $\lesssim$ 10 Myr up to several Gyr old stars,  
the dominant feature of S1 consists of RGB stars, with ages older than $\gtrsim$1-2 Gyr;  a few brighter intermediate-age or old AGB stars
are detected, while stars younger than $\sim$200 Myr are not present in the CMD,  
in agreement with the absence of detected nebular emission  \citep[e.g., ][]{Pustilnik05}. 
The star formation history of S1 was derived through synthetic color magnitude diagrams. The star formation activity at epochs more recent than 
$\sim$500 Myr is compatible with being null, and the large majority of activity occurred at epochs earlier than $\sim$1 Gyr ago. While the data in our hands do not allow us to infer the details of the SFH at epochs earlier than $\sim$1 Gyr ago, the derived total mass of formed stars, $\sim 1.3 \times 10^{6} M_{\odot}$, 
can be considered a quite robust result.

S1 is comparable in terms of  stellar mass to 
the two most classic dSph satellites of the Milky Way,  Draco and Ursa Minor \citep[e.g.][]{McConnachie12}. 
However S1, with a major-axis half-light radius of $R_{e,maj}\simeq$1.9 kpc, is significantly more extended than Draco and Ursa Minor,
with total major axis diameters of $\sim$0.2 kpc. Indeed, from a comparison with the dwarfs in the compilation of  
\cite{McConnachie12}, S1 appears exceptionally extended for its luminosity ($M_r=-9.0 \pm 0.4$), and comparable only to the dSph galaxy And~XIX (see Fig.~6 of  that paper). S1 is also remarkable in terms of its high ellipticity: in the McConnachie compilation, only 4 galaxies out of 110 have $\epsilon\gtrsim0.7$. 

All these properties suggest that S1 has been deeply transformed  under the effect of DDO~68's gravitational potential.
This scenario is supported by N-body simulations only including the collisionless component (stars and dark matter), which are able to reproduce the observed optical properties of  S1 (shape, extension, surface brightness)  with the recent accretion by DDO~68 of a satellite with total (stellar plus dark matter) mass lower by a factor 150 than the total mass of DDO~68  \citep{Annibali16}. 
It is worth noting that the stellar mass ratio between S1 and DDO~68 is  $\approx 1.3\times10^6/1.3\times10^8=1/100$, which is comparable to the mass ratio between satellite and main galaxy in the simulation. Based on estimates of the stellar-to-halo mass relation for dwarf galaxies \citep{Read17}, we would expect  the ratio between the halo masses to be higher than 1/150, but it is possible that the satellite's halo has been largely stripped in the early phases of the interaction with DDO~68.

Existing VLA and GBT HI data  \citep{Cannon14} show modest HI column densities at the position of S1 
($N_{HI}\lesssim10^{20} cm^{-2}$); however, it is not possible to ascertain if any gas  is in fact associated with S1, which 
turns out to be completely embedded into DDO~68's HI emission due to its close proximity to the dwarf galaxy main body. 
Admittedly, the lack of stellar populations younger than $\sim$500 Myr would suggest S1 to be a gas-poor system. 

A possible scenario is that the SF in S1 was quenched some hundred Myrs ago through gas stripping by DDO~68. Nevertheless, it is very 
unlikely that S1 had any significant role in keeping the present-day gas metallicity of DDO~68 so low. The total HI gas mass of DDO~68 was estimated  to be $\sim10^9 M_{\odot}$ by \cite{Cannon14}; even if S1's gas content was a factor ten more massive than its stellar mass of $\sim10^6 M_{\odot}$, its contribution to DDO~68' s gas mass would amount to just 1\%, too small to significantly dilute the metallicity of DDO~68' s interstellar medium. 
Most likely, the culprit of DDO68' s extremely low metallicity is the already merged DDO~68~B system, with a mass of one tenth that of DDO~68's main body.

The case of S1 is particularly relevant in the light of the predictions of the hierarchical galaxy formation process.
Dark-matter only simulations in a  $\Lambda$CDM cosmology predict that substructures persist within haloes down to the resolution limit of the simulations \citep{Diemand08}; hydrodynamical simulations including different sources of feedback \citep[e.g.][]{Wheeler15,Wheeler18} have then attempted to predict the fraction of sub-haloes that hosts in fact  star formation. For instance, according to \cite{Dooley17} we expect a $\sim3\times10^{8}$ M$_{\odot}$ stellar mass galaxy (i.e., with mass not too far from that of DDO~68) to host 1$-$2 satellites with stellar masses above $10^{5} M_{\odot}$.  These predictions are in excellent agreement with the case of DDO~68, which is accreting a ten times smaller body (DDO~68~ B) and another smaller $10^{6} M_{\odot}$  stellar mass system (S1).

\acknowledgments

These data are associated with the HST GO program 14716 (PI: F. Annibali).  
Support for program number 14716 was provided by NASA through 
a grant from the Space Telescope Science Institute, which is operated by the 
Association of Universities for Research in Astronomy under NASA contract NAS5-26555. 
FA, MC, and MT kindly acknowledge funding from INAF PRIN-SKA-2017 program 1.05.01.88.04. 
We thank the anonymous referee for her/his useful comments and suggestions.

%

\vspace{5mm}
\facilities{HST(ACS and WFC3), LBT(LBC)} 

\software{Drizzlepac \citep{gonz+12}  
          Dolphot \citep{dolp00}, 
          }

\end{document}